\documentclass[aps,prb,twocolumn,showpacs]{revtex4-1}
\usepackage{euscript}
\usepackage{amssymb}
\usepackage{amsmath}
\usepackage{graphicx}

\begin{document}

\title{Photoinduced polarons in polymers. Time-resolved ESR
analysis of polaron pairs in polymer:fullerene blends.}

\author{A. I. Shushin}
\affiliation{Institute of Chemical Physics, Russian Academy of
Sciences, 119991, GSP-1, Kosygin St. 4, Moscow, Russia}

\begin{abstract}
The work concerns the analysis of experimental time-resolved ESR
spectra in photoexcited polymer:fullerene blend, consisting of
poly(3-hexilthiophene) and fullerene [6,6]-phenyl ${\rm
C}_{61}$-butyric acid methyl ester (at low temperature $T = 100
\,{\rm K}$). The spectra are assumed to be determined by
spin-coherent pairs of charged polarons (P${}^{^{_+}}\!$ and
P${}^{^{_-}}\!\!$) generated in the singlet state. The analysis is
made within simple model of a set of first order processes, in which
P${}^{^{_+}}\!$P${}^{^{_-}}\!\!$-pair spin evolution is described by
the stochastic Liouville equation, allowing for fairly accurate
description of experimental results. Simple analytical
interpretation of obtained numerical results demonstrates that trESR
spectra can be represented as a superposition of antiphase and CIDEP
contributions together with the conventional thermal one. These
contributions are shown to change their signs with the increase of
time in agreement with experimental observations.
\end{abstract}


\maketitle

\section{Introduction}

Time-resolved ESR (trESR) spectra of pairs of doublet (D) particles
and, in particular, spin-correlated pairs have been studied very
actively both experimentally
\cite{Hore1,Forb1,Hore2,Forb2,Hore3,Forb3,Forb4,Lev1,Kub1,Forb5,
Hir1,Hir2,Forb6,Hore0} and theoretically.
\cite{Hore0,Hore1,Forb1,Hore2,Shu1,Shu2,Ped1,Forb7} The analysis of
these spectra is known to give important information on
characteristic properties of spin-dependent interactions between
particles in pairs (Zeeman, spin-exchange, spin dipole-dipole, and
hyperfine, etc.) as well as spin relaxation and spin-dependent
recombination rates.

The interference of effects of spin-dependent interactions and
relaxation/recombination leads to well known contributions to trESR
spectra: antiphase structure (APS), chemically induced dynamic
electron polarization (CIDEP) both multiplet and net, and some other
contrbutions. The manifestation of them is found in a large number
of photochemical condensed phase D-D reactions, charge separation in
photosynthetic systems, etc.\cite{Hore0}

Close attention has, recently, been attracted to the trESR
investigation of photoinduced charge transfer processes in solids
(organic semiconductors, polymer blends, etc.),
\cite{Heeg1,Pope,Green1,Dyak1} which determine the efficiency of
generation of charged D-particles, i.e. P$^{^{\!_\pm}\!}$-polarons.
Great interest in studies of charge transfer processes is inspired
by possible practical applications in spintronics\cite{Zut} and
solar cells.\cite{Pope}

Some recent experiments concern trESR investigation of photoexcited
polymer:fullerene blend, consisting of poly(3-hexilthiophene) (P3HT)
and fullerene [6,6]-phenyl ${\rm C}_{61}$-butyric acid methyl ester
(PCMB), at low temperature $T = 100 \,{\rm K}$.\cite{Beh1} The
observed trESR spectra indicate the generation of geminate
spin-correlated doublet-doublet pairs in the initial singlet state,
which are suggested to be the pairs of polarons P${}^{^{_+}}_{}$ (in
P3HT) and P${}^{^{_-}}_{}$ (in PCMB),\cite{Beh1} called hereafter
P${}^{^{\!_+}}_{}\!$P${}^{^{\!_-}}_{}\!\!$-pairs. The shape of trESR
spectra, observed at times  $1\, {\rm \mu s} \lesssim t \lesssim 30
\, {\rm \mu s}$, is found to significantly change with time: at
small times $\sim 1\, {\rm \mu s} $ the spectrum is of the APS
shape, which at later times is changed by more complicated one still
with the APS-type contribution, though of the sign opposite to that
observed at $t \sim 1\, {\rm \mu s} $.

We demonstrate that above-discussed trESR spectra can interpreted
within simple exponential model of
P${}^{^{\!_+}}_{}\!$P${}^{^{\!_-}}_{}\!\!$-pair
evolution,\cite{Shu2,Shu3} taking into consideration the Zeeman and
electron spin-exchange interactions, spin-exchange-induced and the
Bloch-type intrapolaron spin relaxation, spin-selective
P${}^{^{\!_+}}_{}\!$P${}^{^{\!_-}}_{}\!\!$ recombination (in the
singlet state of the pair), and dissociation. The
P${}^{^{\!_+}}_{}\!$P${}^{^{\!_-}}_{}\!\!$ spin evolution is
described by the stochastic Liouville equation (SLE),\cite{Muu}
which is solved numerically, though the most important numerical
predictions are clarified with the simple approximate analytical
treatment. The results of this combined numerical/analytical
analysis shows that the proposed simple model can quite
satisfactorily describe the time evolution of experimental trESR
spectra of P${}^{^{\!_+}}_{}\!$P${}^{^{\!_-}}_{}\!\!$ pairs under
study.

The proposed analysis has shown that at all times observed spectra
can be represented as a superposition of APS and CIDEP contributions
superimposed with conventional thermal
P${}^{^{\!_+}}_{}\!$P${}^{^{\!_-}}_{}\!\!$ one. The APS and CIDEP
contributions are found to change their sign at long times because
of the effect of spin-selective reactivity and spin population
relaxation on spin evolution of
P${}^{^{\!_+}}_{}\!$P${}^{^{\!_-}}_{}\!\!$ pairs. Naturally, the
amplitudes of obtained APS and CIDEP contributions decrease with the
increase of times but remain quite distinguishable against the
background of the thermal ESR spectrum even at long times $t \gtrsim
10\, {\rm \mu s}$.

\section{General formulation}

The trESR spectra of
P${}^{^{\!_+}}_{}\!$P${}^{^{\!_-}}_{}\!\!$-pairs are determined by
the spin evolution of interacting P${}^{^{\!_\pm}}_{}\!$ polarons.
In this work the spin evolution is described by the
P${}^{^{\!_+}}_{}\!$P${}^{^{\!_-}}_{}\!\!$ spin density matrix $\rho
(t)$, satisfying the SLE. The characteristic features of the
evolution essentially depend on the form of the SLE, which, in turn,
depends on the model of the process under study. In our work we
consider the P${}^{^{\!_+}}_{}\!$P${}^{^{\!_-}}_{}\!$ spin evolution
in the simple exponential model.\cite{Shu2} In this model SLE is
written as ($\hbar = 1$)
\begin{equation} \label{gform1}
\dot \rho = -\hat L \rho \;\;\mbox{with}
\;\; \hat L = i\hat H + \hat W_{\!r} + \hat W_{\!e} + W_d + \hat K_s.
\end{equation}
Here $\hat L$ is the superoperator [operator in the Liouville space
(vide infra)], in which $\hat H$ is the superoperator representation
of the commutator of the P${}^{^{\!_+}}_{}\!$P${}^{^{\!_-}}_{}\!\!$
spin-Hamiltonian $H$, defined by $\hat H \rho = H\rho - \rho H$,
$\hat W_{\!r}$ is the superoperator of intraradical spin relaxation,
$\hat W_{\!e}$ is the superoperator of spin-exchange-induced
relaxation rate, $W_d$ is the rate of irreversible PP dissociation,
and $\hat K_s$ is the spin-selective-recombination rate
superoperator.

Hereafter we will use the term "superoperator" and the notation
$\hat {A}$ (with hat) for any operator $A$ in the Liouville space
(the space, in which the density matrix is represented as the
vector), in order to distinguish this operator from those in the
Hilbert space.

In our analysis we consider the pairs of P${}^{^{\!_\pm}}_{}\!$
polarons, with different electron spin $g$-factors, $g_{_+}^{}$ and
$g_{_-}^{}$, and exchange interaction $J_e$, in the magnetic field
$B$. The P${}^{^{\!_+}}_{}\!$P${}^{^{\!_-}}_{}\!\!$ spin-Hamiltonian
is represented in a simple form
\begin{equation} \label{gform2}
H = {\omega}S_{_z} + Q (S_{_{\!+_z}}\!\! - S_{_{\!-_z}})
- J_e^{} (S^2\! - 1).
\end{equation}
In this formula
\begin{equation} \label{gform2a}
\omega = \mbox{$\frac{1}{2}$}(g_{_+}\! + g_{_-\!})\beta B
\;\; \mbox{and} \;\;
Q  = \mbox{$\frac{1}{2}$}(g_{_+}\! - g_{_-\!})\beta B;
\end{equation}
$S_{\nu_z},\, (\nu = \pm),\,$ are the $z$ components (along the
magnetic field direction) of the P${}^{^{\!_\pm}}_{}\!$-spin
operators with eigenvectors, corresponding to the projections $+1/2$
and $-1/2$, denoted as $|\!\!\uparrow\rangle_{\!_\nu}$ and
$|\!\!\downarrow\rangle_{\!_\nu}$, respectively; ${\bf S} = {\bf
S}_{_+} + {\bf S}_{_-}$ is the operator of the total spin and
$S_{\!_z}$ is its $z$ component. First two terms in eq.
(\ref{gform2}) represent the Zeeman interaction (of electron spins)
with frequences $\omega_{\nu}^{} = g_{\nu}^{} \beta B, \; (\nu =
\pm)$, and the third term describes the spin exchange interaction.

The intraradical spin relaxation is modeled within the simple
Bloch-type approach with relaxation supermatrix
\begin{equation} \label{gform3}
\hat W_{\!r}^{} = \hat W_{\!_+ }+ \hat W_{\!_-},
\end{equation}
in which
\begin{eqnarray} \label{gform4}
\hat W_{\!\nu}^{} &=&
w_{_{^\nu}}^{_d}(p_{\nu}^{}|\!\!\uparrow\uparrow\rangle_{\!_\nu} -
|\!\!\downarrow\downarrow\rangle_{\!_\nu}\!)
({}_{_{\!\nu}\!\!}\langle \uparrow\uparrow\!\!| -
{}_{_\nu\!\!}\langle \downarrow\downarrow\!\!|_{\!_{}}) \nonumber \\
&& +  w_{_{^\nu}}^{_n}(\,|\!\!\uparrow\downarrow\rangle_{\!_\nu}\!
{}_{_\nu\!\!}\langle \downarrow\uparrow\!\!| +
|\!\!\downarrow\uparrow\rangle_{\!_\nu} \!
{}_{_\nu\!\!}\langle \uparrow\downarrow\!\!|\,), \quad (\nu = \pm),\quad
\end{eqnarray}
are the supermatices of relaxation in
P${}^{^{\!_\pm}}_{}\!$-polarons, represented in the Liouville space
in the bases $|\mu\mu'\rangle_{\!_\nu}\! = |\mu\rangle_{\!_\nu}\!
{}_{_\nu\!\!}\langle \mu'|, \: (\mu,\mu' =
\,\uparrow,\downarrow)$.\cite{Shu4}

The terms in first and second lines of formula (\ref{gform4})
describe the population and phase relaxation ($\sim
w_{_{^\nu}}^{_d}$ and $\sim w_{_{^\nu}}^{_n}$), respectively. The
parameters $p_{\nu}^{} = e^{-\omega_{\nu}^{}/(k_{B}T)}$ are the
Boltzmann factors, controlling the difference of thermal populations
of the Zeeman sublevels. Note that in the considered experimental
conditions the parameters $p_{\nu}^{}$ is very close to unity
because $\omega_{\nu}^{}/(k_{B}T) \sim 10^{-3}$.

So far we have discussed the spin states and corresponding operators
for single P${}^{^{\!_\pm}}_{}\!$ particles. For the analysis of
P${}^{^{\!_+}}_{}\!$P${}^{^{\!_-}}_{}\!\!$-pair spin evolution,
however, we need the proper basis of pair spin states. The most
convenient are the basis of states $| j \rangle,\:(j = 1-4),$ of
separate P${}^{^{\!_\pm}}_{}\!$ particles and the basis of
eigenstates of the total spin ${\bf S}$ ($S\!-\!T$-basis)
$|Z\rangle, \:(Z = S, T_{_{0,\pm}})$. In terms of
$(\,|\!\!\uparrow\rangle_{\!_\nu},
|\!\!\downarrow\rangle_{\!_\nu}\!)$-states of
P${}^{^{\!_\pm}}_{}\!$-particles both sets of basis states are
represented as
\begin{subequations}
\begin{eqnarray}
{}_{\!}|\!\!\uparrow\rangle_{\!_+\!}|\!\!\uparrow\rangle_{\!_-\!\!}\!&=&
|1\rangle_{\!} = |T_{\!_+\!} \rangle; \;\;
{}_{\!}|\!\!\uparrow\rangle_{\!_+\!}|\!\!\downarrow\rangle_{\!_-\!\!}\! =
|2\rangle_{\!\!} =
\!\mbox{$\frac{1}{\sqrt{2}}$}(|T_{_0\!}\rangle\! + \!|S_{} \rangle);\quad
\label{gform4a}\\
{}_{\!}|\!\!\downarrow\rangle_{\!_+\!}|\!\!\uparrow\rangle_{\!_-\!\!}\!&=&
|3\rangle_{\!\!} =
\!\mbox{$\frac{1}{\sqrt{2}}$}(|T_{_0\!}\rangle\! - \!|S_{} \rangle);\;\;
{}_{\!}|\!\!\downarrow\rangle_{\!_+\!}|\!\!\downarrow\rangle_{\!_-\!\!}
= |4\rangle_{\!\!} =|T_{\!_-\!} \rangle.\quad\label{gform4b}
\end{eqnarray}
\end{subequations}

In particular, the recombination rate superoperator $\hat K_{s}$ is
represented in terms of the operator $P_{\!s} = |S \rangle \langle S
|$ of projection on the singlet $|S \rangle$ state:
\begin{equation} \label{gform5}
\hat K_{s} = \mbox{$\frac{1}{2}$}k_s \hat P_{\!s} \;\;\mbox{with}\; \;
\hat P_{\!s}\rho = P_{\!s}\rho + \rho P_{\!s}.
\end{equation}

The superoperator $\hat W_{\!e}^{}$ of the rate of
spin-exchange-induced relaxation can also be written in terms of
this operator:\cite{Shu2}
\begin{equation} \label{gform5a}
\hat W_{\!e}^{}\rho = w_e^{} (P_{\!s}\rho + \rho P_{\!s}
- 2 P_{\!s}\rho P_{\!s}).
\end{equation}

To obtain the spin density matrix $\rho$ [from the SLE
(\ref{gform1})] one needs to specify initial condition
$\rho(t=0)=\rho_{i}^{}$. In our work we will mainly consider the
case of initial creation of PPs in the singlet ($\rho_{\!_S}^{}$) or
the thermalized triplet $(\rho_{\!_T}^{})$ state:
\begin{equation} \label{gform5b}
\rho_{i}^{} = \rho_{\!_S}^{}\! = |S \rangle \langle S |
\;\;\mbox{and} \;\; \rho_{i}^{} = \rho_{\!_T}^{}\! =
\mbox{$\frac{1}{3}$}
\sum\nolimits_{\!\!_{\nu=0,\pm}^{}}\!\!|T_{\!\nu}^{}
\rangle\langle T_{\!\nu}^{}|,
\end{equation}
respectively.

The specific features of the above-described
P${}^{^{\!_+}}_{}\!$P${}^{^{\!_-}}_{}\!$ spin dynamics essentially
manifest themselves in the trESR spectrum, formed during the time of
P${}^{^{\!_+}}_{}\!$P${}^{^{\!_-}}_{}\!$ spin evolution in the
presence of the microwave field $B_1^{}$, assumed to rotate with the
frequency $\omega_0^{}$ around the magnetic field ${\bf B}$ (around
the axis $z$). The spectrum is determined by the
P${}^{^{\!_+}}_{}\!$P${}^{^{\!_-}}_{}\!$ spin evolution, governed by
the Liouville superoperator $\hat L_{\!_\omega}$. In the fame of
reference ($x', y',z$), rotating with $B_1^{}$, this superoperator
is written as
\begin{equation} \label{gform6}
\hat L_{\!_\omega} = i(\hat H_{\!_\omega} +
\hat H_1^{}) + \hat W_{\!r} + \hat W_{\!e} + W_d + \hat K_s.
\end{equation}
In this formula $\hat H_{\!_\omega}$ is the superoperator
representation of the commutator of the spin-Hamiltonian
\begin{equation} \label{gform6a}
H_{\!_\omega} = H -\omega_0^{} S_{z}^{} \;\; \mbox{and} \;\;
H_1^{} = \omega_1^{} S_{x'}^{},
\end{equation}
in which $\omega_1^{} \approx \frac{1}{2} (g_{a}^{}\! + g_{b}^{})
\beta B_1^{}$. In terms of the superoperator $\hat L_{\!_\omega}$
the trESR spectrum $I_t^{}(\omega)$ can be expressed as
\begin{equation} \label{gform7}
I_{t}^{}(\omega) = \!\int_{_0}^{^\infty}\!\!\!\!d\tau f(\tau)
{\rm Tr} [S_{y'}^{}(e^{{-\hat L_{\!_\omega}\!\tau}}\!\rho_t^{})]
= {\rm Tr} [S_{y'}^{}\!(\hat F_{\!_\omega}\rho_t^{})],\!
\end{equation}
where $f(\tau) = w_g^{}e^{-w_{g}^{}\tau}$ is the characteristic
window function\cite{Forb7} of the ESR spectrometer, with small time
$w_g^{-1} \ll t$, $\rho_t^{}\equiv \rho (t)$ is the
P${}^{^{\!_+}}_{}\!$P${}^{^{\!_-}}_{}\!$ spin-density matrix at the
moment $t$ of measurement, and
\begin{equation} \label{gform8}
\hat F_{\!_\omega} = w_g^{}(w_g^{} + \hat L_{\!_\omega})^{-1}_{}
\end{equation}
is the spectrum-shape matrix.

\section{Numerical description}

In general, in the considered model trESR spectra can be evaluated
only numerically. Here we compare numerically calculated spectra
with the experimental ones.

In our numerical treatment of trESR spectra we use the parameters of
the spin-Hamiltonian close to those obtained from
experiments.\cite{Beh1} Moreover, the considered model
(\ref{gform2}), taking into account only the isotropic Zeeman and
exchange interactions, is actually based on the experimental results
and suggestions of the work.\cite{Beh1}

The anisotropic Zeeman and the hyperfine interactions are neglected,
though the effect of these two types of interactions is taken into
consideration, in terms of spin relaxation, which is represented by
the matrix $\hat W_{\!r}^{}$ (\ref{gform3}) (resulted from
polaron-jump-induced fluctuations of these interactions).

Detailed analysis of experimental spectra shows that anisotropic
Zeeman and hyperfine interactions lead also to some inhomogeneous
broadening of lines. In reality, however, it is impossible to
accurately describe this broadening, because of not very high
accuracy of experimental spectra. In such a case it looks quite
appropriate to approximate the effect of inhomogeneous broadening
with that of homogeneous one, by properly adjusting the values of
rates in the relaxation matrix $\hat W_{\!r}^{}$.

The inhomogeneous broadening can also result in some (small) change
of frequencies of lines. Taking into account this effect we will
slightly adjust the (effective) Zeeman frequences $\omega_{\nu}^{},
\,(\nu =\pm)$ in the spin-Hamiltonian (\ref{gform2}) to the average
experimental ones (see below).

In this section we show fairly good accuracy of the model, proposed
and discussed in Sec. II. Despite evident limitations of this
semiquantitative  model, its accuracy turns out to be high enough to
demonstrate the most important specific features of the time
evolution of experimental trESR spectra of
P${}^{^{\!_+}}_{}\!$P${}^{^{\!_-}}_{}\!\!$-pairs.

The model contains the parameters of two types: (a) specifying the
measurement conditions and (b) describing
P${}^{^{\!_+}}_{}\!$P${}^{^{\!_-}}_{}\!$ pair under study.

The parameters of the first type (a) are the microwave field
$\omega_1^{}$ (in frequency unit) and characteristic inverse time
$w_g^{}$ of the window function $f(t)$ [see eq. (\ref{gform7})]. The
field $\omega_1^{}$ is considered to be weak $\omega_1^{} \ll
w_{\nu}^{n}, k_s$ (corresponding to experimental conditions
\cite{Beh1}), for which trESR spectra $I_t^{}(\omega) \sim
\omega_1^{2}$ and the shape of these spectra are independent of
$\omega_1^{}$. As for the parameter $w_g^{}$, its value satisfies
the relation $w_{\nu}^{_d} \ll w_g^{} \ll w_{\nu}^{_n}, \,(\nu =
\pm)$, which on the one hand, ensures that measurements do not lead
to additional broadening (due to uncertainty principle) and on the
other hand, the important kinetic features of the trESR signal
generation are well resolved (as was realized in the experimental
analysis\cite{Beh1}).

The most important parameter of the second type (b) is $Q =
\frac{1}{2} (\omega_{_+}\! - \omega_{_-}) = 3.25\cdot 10^{7} \, {\rm
s}^{-1}$. This value is slightly larger ($\approx 12\%$), than that
$Q_{ g}^{} \approx \frac{1}{4}(g_{_+}\! - g_{_-})\omega_0$,
corresponding to $\Delta g = g_{_+}\! - g_{_-}= 1.9\cdot 10^{-3}$
(obtained in ref. [24] from cwESR spectra of P${}^{^{\!_\pm}}_{}\!$
polarons).

All other parameters of the model are considered as adjustable in
fitting of experimental data. Some relations between relaxation
rates are, however, evident without numerical analysis.

In particular, experimental results clearly show that widths of
trESR lines of P${}^{^{\!_+}}_{}\!$P${}^{^{\!_-}}_{}\!$ pairs are
fairly large, corresponding to dephasing rates $w_{_{^\nu}}^{_n}$
comparable with the difference of Zeeman frequences $Q$:
$w_{_{^\nu}}^{_n} \lesssim Q \sim (2-4) \cdot 10^{7} \,{\rm s}^{-1},
\, (\nu = \pm)$. At the same time, the population relaxation rates
$w_{_{^\nu}}^{_d}$ are much smaller than the dephasing rates
$w_{_{^\nu}}^{_d}/w_{_{^\nu}}^{_n} < 0.1$, as it follows from the
characteristic time of changing of the experimental trESR spectrum
$\tau_{\!_{ESR}} \sim 10^{-6}\, {\rm s}$ (see Fig. 1). The obtained
estimation $w_{_{^\nu}}^{_d} \ll w_{_{^\nu}}^{_n}$ allows for the
proper choice of the inverse width of the window function $w_g^{}$
mentioned above.

\begin{figure}
\setlength{\unitlength}{1cm}
\includegraphics[height=8.5cm,width=8cm]{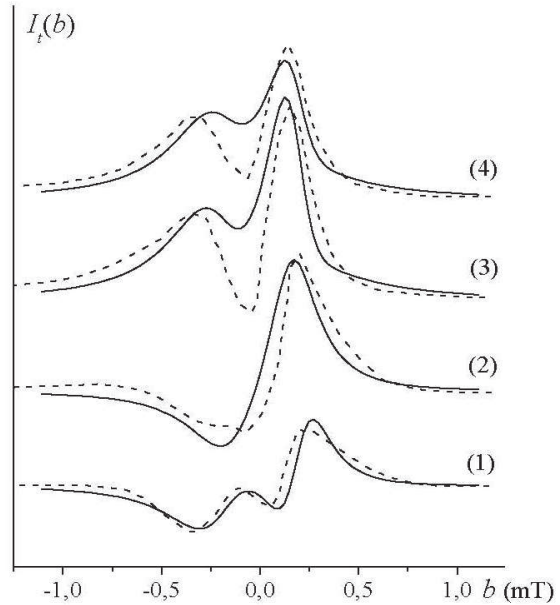}
\caption{Comparison of experimental (dashed lines) and theoretical (full
lines) ESR spectra of P${}^{^{\!_+}}_{}\!$P${}^{^{\!_-}}_{}\!$ pairs
$I_t^{}(b)$ (in arbitrary units) [with $b = B-B_0^{}$ defined in eq.
(\ref{numres2})] for four times $t \: ({\rm \mu s})$: (1) $ 0.8\, $,
(2) $ 1.4\,$, (3) $ 3.5\,$, (4) $5.8$. Other parameters used are $Q
= 3.25 \cdot 10^7 \, {\rm s}^{-1}, \, $ $J_e^{} = 7.3 \cdot 10^5 \,
{\rm s}^{-1}, \, $ $k_s^{} = 6.0 \cdot 10^5 \, {\rm s}^{-1}, \,$ and
$w_g^{} = 3.2 \cdot 10^6 \, {\rm s}^{-1}, \,$ as well as relaxation
rates $w_{^{_+}}^{_n} = 4.2 \cdot 10^7 \, {\rm s}^{-1}, \, $
$w_{^{_-}}^{_n} = 2.5 \cdot 10^7 \, {\rm s}^{-1}, \, $
$w_{^{_+}}^{_d} = 2.0 \cdot 10^5 \, {\rm s}^{-1}, \, $
$w_{^{_-}}^{_d} = 6.5 \cdot 10^5 \, {\rm s}^{-1}.$}
\end{figure}

Special assumption is made on the absolute value of the
spin-exchange interaction $J_e^{}$. Analysis shows that the best
agreement between experimental and theoretical trESR spectra is
obtained for small $|J_e^{}| \ll |Q|, w_{_{^\nu}}^{_n}$ in
accordance with the suggestion of the work.\cite{Beh1} In this limit
the trESR shape does not strongly depend of $J_e^{}$.

In our calculations we neglect the effect of the dissociation of
P${}^{^{\!_+}}_{}\!$P${}^{^{\!_-}}_{}\!$ pair by taking $W_{\!d}^{}
= 0$, because the decay manifests itself in trESR spectrum only at
very long times ($t > 10 {\rm \mu s}$), which, first, are not of
special interest for the analysis of spin-correlated
P${}^{^{\!_+}}_{}\!$P${}^{^{\!_-}}_{}\!$ pairs and, second, the
experimental ESR spectra at these times are not accurate enough for
unambiguous interpretation. Though, it is worth noting that some
contribution of the dissociation process improves agreement between
theoretical and experimental trESR spectra at $t > 10 \, {\rm \mu
s}$ (as it follows from the numerical investigation).

The proposed analysis does not allow to clearly distinguish the
effect of the spin-exchange induced relaxation $\hat W_{\!e}^{}$
against that of relaxation in free polarons $\hat W_{\!_\pm}$,
especially taking into account that $\hat W_{\!e}^{}$-effect can
hardly be distinguished from that of reactivity $\hat K_s^{}$ in the
system under study. For this reason, the spin-exchange induced
relaxation is also neglected ($\hat W_{\!e}^{} = 0$).

As for the reactivity, according to the analysis, good agreement
between experimental and theoretical results is found for not very
large values $k_s^{} \lesssim w_{_{\!^\pm}}^{_d}$.

Figure 1 displays the comparison of experimental and numerically
calculated trESR spectra $I_t^{} (b)$, in which
\begin{equation} \label{numres2}
b = B - B_0^{}, \;\;\mbox{where}\;\; B_0^{} = \omega_0^{}/(\bar g \beta)
\end{equation}
with $\bar g = \frac{1}{2}(g_{_{+}\!}\! + g_{_{-}})$, for four
values of time delays: $\:0.8, \: 1.4,\: 3.5,\,$ and $\,5.8 \: ({\rm
\mu s})$.

The spectra (in Fig. 1) are obtained for the singlet ($|S\rangle$)
initial state\cite{Beh1} of P${}^{^{\!_+}}_{}\!$P${}^{^{\!_-}}_{}\!$
pairs, using eq. (\ref{gform7}) with $\omega(B(b))$ defined in eq.
(\ref{gform2a}).

It is seen that the proposed model allows one to reproduce
experimental spectra fairly accurately, particularly taking into
consideration that the accuracy of measured spectra is not very
high.

The observed bits between lines of experimental spectra at $t =
3.5\,$ and $\,5.8 \; ({\rm \mu s})$ are somewhat deeper than those,
obtained in the considered model. This discrepancy can be caused by
the approximation of partially inhomogeneous broadening of the trESR
spectra by homogeneous one, which results in relatively broad wings
of lines. The fact is that the homogeneous mechanism predicts
Lorentzian line shape with slowly decreasing wings while the
inhomogeneous mechanism usually yields Gaussian one with very
rapidly decreasing wings. The slower decrease of Lorentzian lines
can certainly lead to less deep bits between Lorentzian lines than
those between Gaussian ones.

Concluding discussion we would like to draw attention to the fact
that E/A-distortion of trESR spectra, observed at small times
$\,0.8\,$ and $ \, 1.4\, ({\rm \mu s})$ is changed by the A/E-one at
longer times $\,3.5\,$ and $\,5.8 \: ({\rm \mu s})$. Of course, at
long times relative distortion amplitudes are fairly small, but
still quite pronounced and distinguishable. It is seen that the
calculation within the proposed model correctly reproduces the
distortion behavior.

\section{Analytical analysis of ESR spectra}

Detailed analysis shows that the shape of trESR spectra, predicted
by the proposed model, can markedly be changed with the change of
parameters of the model. There are, however, some important general
properties, which essentially determine the shape. Here we discuss
them both analytically and numerically.

\subsection{APS contribution}

One of the most well known manifestations of the spin-correlation of
P${}^{^{\!_+}}_{}\!$P${}^{^{\!_-}}_{}\!$ population is the
APS.\cite{Hore0} To demonstrate important features of the APS
contribution we will discuss two simple variants of spin
correlation, corresponding to the population of the singlet
$|S\rangle$-state and isotropic triplet $|T\rangle$-state [see eq.
(\ref{gform5b})].

The most important specific features of the APS can be illustrated
with the simplest model, based on the analysis of
$\omega_1^{}$-induced transitions between spin states of the
Hamiltonian $H_{\omega}^{}$ (\ref{gform2}), using simple
approach\cite{Hore0}
\begin{equation} \label{spepr1}
I_{\!_{A}}^{^{_{_{\rm X}}}}(\omega) = \sum\nolimits_{_{jk}}
\!\!P_{\!_{{jk}}}^{^{_{_{\rm X}}}}\! \Phi (\omega -\omega_{\!jk}^{}),
\;({\rm X = S, T}),
\end{equation}
where  $\Phi (\omega)=1/({\bar w}^2 + \omega^2)$ is the shape
function with the characteristic width $\bar w \sim w_{\pm}^{n}$. In
eq. (\ref{spepr1}) the sum is taken over the pairs of states ($j,k =
1-4$) of the spin-Hamiltonian $H_{\omega}^{}$, involved in ESR
($H_1^{}$-induced) transitions, and superscript ${\rm X = S,T}$ of
parameters $P_{\!_{{jk}}}^{^{_{_{\rm X}}}}$ denotes
P${}^{^{\!_+}}_{}\!$P${}^{^{\!_-}}_{}\!\!$-pair spin state, defined
in eq. (\ref{gform5b}). Amplitudes of lines are determined by
transitions intensities $P_{\!_{{jk}}}^{^{_{_{\rm X}}}} = P_0^{}
(p_{j}^{_{\rm X}} - p_{k}^{_{\rm X}}), \, ({\rm X = S, T})$, in
which $p_{j}^{_{\rm X}}, p_{k}^{_{\rm X}}$ are the populations of
the initial ($j$) and final ($k$) states, and $P_0^{}$ is a constant
parameter.

To further simplify the problem we consider the limit of weak
exchange interaction $|J_e^{}| \ll |\omega_{\!_+}\! -
\omega_{\!_-}|, \bar w$. In this limit, i.e. for the small parameter
$\xi_e = |J_e^{}|/|Q| \ll 1$, the APS can be interpreted, neglecting
the effect of the exchange interaction on eigenstates of the
spin-Hamiltonian $H_{\omega}^{}$, i.e. assuming them to coincide
with eigenstates $|j\rangle, \: (j=1-4),$
[(\ref{gform4a})-(\ref{gform4b})]. Within this approximation the APS
results from the change of corresponding eigenenergies:
\begin{equation} \label{spepr2}
\epsilon_1^{} \approx \Delta\omega - J_e^{}; \;
\epsilon_2^{} \approx Q; \;
\epsilon_3^{} \approx -Q;\;
\epsilon_4^{} \approx -\Delta\omega - J_e^{},
\end{equation}
with $\Delta\omega = \omega - \omega_0^{}$ and $Q =
\mbox{$\frac{1}{2}$}(\omega_{\!_+}\! - \omega_{\!_-})$, for which we
get the following transition frequences $\omega_{\!jk}^{}$:
$\omega_{\!12}^{} \approx \omega_{\!+}^{}\! - J_e^{}, \:
\omega_{\!34}^{} \approx \omega_{\!+}^{}\! + J_e^{}, \:
\omega_{\!13}^{} \approx \omega_{\!-}^{}\! - J_e^{}, \:
\omega_{\!24}^{} \approx \omega_{\!-}^{}\! + J_e^{}$. As to
populations $p_{j}^{_{\rm X}}$, they depend on the spin state X: for
the singlet state ($S = 0$) $p_{1,4}^{_{\rm S}} = 0, \:
p_{2,3}^{_{\rm S}} = 1/2,$ and for the triplet state ($S = 1$)
$p_{1,4}^{_{\rm T}} = 1/3, \: p_{2,3}^{_{\rm T}} = 1/6$.
Substitution of these values into eq. (\ref{spepr1}) yields for
${\rm X = S, T}$\cite{Shu1}
\begin{equation} \label{spepr3}
I_{\!_{A}}^{^{_{_{\rm S, T}}}}(\omega)
\sim  (-1)^{^{_{\!_{S}}}}\! J_e^{}\,
[\Phi^{\prime} (\omega_{_+}\!\!-\omega) +
\Phi^{\prime} (\omega_{_-}\!\!-\omega)],
\end{equation}
where $\Phi^{\prime} (\omega) = d\Phi (\omega)/d\omega = -2
\omega/({\bar w}^2 + \omega^2)^{2}$.

\subsection{$ST_0^{}$-CIDEP contribution}

Another important contribution, which essentially determines the
shape of trESR spectra of
P${}^{^{\!_+}}_{}\!$P${}^{^{\!_-}}_{}\!$-pairs, is the multiplet
CIDEP.\cite{Muu} Specific features of the kinetics of the multiplet
CIDEP generation can be analyzed within the simple, but quite
realistic limit of weak spin exchange interaction and dephasing
faster than the population relaxation and recombination, i.e. for
$w_{^{\!_\pm}}^{_{n}} \gg w_{^{\!_\pm}}^{_{d}}, k_s^{}$.

In the natural case of large magnetic field $B$ and for not very
long times $t < 1/w_{^{\!_\pm}}^{_d}, 1/k_s^{}$ the multiplet CIDEP
generation is described by the $ST_0^{}$-mechanism,\cite{Muu}
neglecting population relaxation and recombination (see Sec. IV.A).
The $ST_0^{}$-CIDEP generation kinetics is described by the SLE
(\ref{gform1}) for the reduced spin-density matrix $ \rho_{{r}}^{}
(t)$ in the subspace $(|S\rangle,|T_0^{}\rangle) \equiv
(|2\rangle,|3\rangle)$, which in the vector form is represented as $
\rho_{{r}}^{} = [\rho_{_{22}}, \rho_{_{23}}, \rho_{_{32}},
\rho_{_{33}}]_{}^{^{_{\sf T}}}$. For $\tilde {\rho}_{{r}\!}^{\!}
(\epsilon)\! = \! \int_0^{\infty}\!dt \rho_{\!{r}}^{}(t)
e^{-\epsilon t}\!$ this SLE is written as
\begin{equation} \label{spepr5}
(\epsilon + \hat L_{r}^{})\tilde {\rho}_{{r}}^{} = \rho_i^{_0}
\end{equation}
with
\begin{equation} \label{spepr6}
\hat L_r^{} = \left[
\begin{array}{cccc}
0   &   iJ_e^{}     &  \!-iJ_e^{}       &        \,\, 0   \\
iJ_e^{}  &   \Omega &   \,0        &         \!-iJ_e^{}   \\
\!-iJ_e^{}\, &   0      & \,\,\Omega^{*} &  \,\, iJ_e^{}  \\
0   &   \!\!-iJ_e^{}    &  \,iJ_e^{}        &    \,\, 0
\end{array}
\,\right] \qquad \left[
\begin{array}{c}
\tilde \rho_{_{22}} \\
\tilde \rho_{_{23}} \\
\tilde \rho_{_{32}} \\
\tilde \rho_{_{33}}
\end{array}
\right].
\end{equation}
Here $\Omega = w_{\!n}^{} + 2iQ$ with $w_{\!n}^{} = w_{^{\!_+}}^{_n}
+ w_{^{\!_-}}^{_n}$, and $\rho_i^{_0} = \frac{1}{2}
[1,(-1)^{^{_{1+S}}},(-1)^{^{_{1+S}}},1]_{}^{^{_{\sf T}}}$ is the
initial spin-density matrix for singlet ($S = 0$) and triplet ($S =
1$) initial state.

Equation (\ref{spepr5}) can be solved analytically in the studied
case of fast dephasing and small $|J_e^{}| \ll w_{^{\!_\pm}}^{_n}$.
The solution leads to the following expression for the Laplace
transform of the multiplet CIDEP $\tilde
P_{\!_{^{^\pm}}}^{^{\!_M\!}}\! (\epsilon) = -2 \langle S_{_{\pm_z}}
 \rangle_{\!\epsilon}^{}$:
\begin{eqnarray} \label{spepr7}
\tilde P_{\!_{^{^+}}}^{^{\!_M\!}}\!(\epsilon) &=& -
\tilde P_{\!_{^{^-}}}^{^{\!_M\!}}\!(\epsilon) =
2[\tilde \rho_{_{33}}(\epsilon) - \tilde \rho_{_{22}}(\epsilon)]\nonumber\\
&\approx& 4\sigma_{\!_{^S}\!}J_e^{}Q
\Big\{\epsilon[(\epsilon_{\!\!} + {\!}_{}w_{\!n}^{})^2\! + 4Q^2]
+ 4J_e^{2}\!(\epsilon_{\!\!} +
{\!}_{}w_{\!n}^{})\!\Big\}_{}^{\!_{^{-1}}}\!\!, \quad\;\;
\end{eqnarray}
where $\sigma_{\!_{^S}\!} = (-1)^{\!^{_S}}\!$ with $S = 0$ and $S =
1$ for the singlet and triplet initial condition, $\rho_i^{_0} = |S
\rangle \langle S|$ and $\rho_i^{_0} = |T_0^{} \rangle \langle
T_0^{}|$  respectively. The inverse Laplace transformation of the
expression (\ref{spepr7}) yields for small $|J_e^{}| \ll w_{\!n}^{}$
\begin{equation} \label{spepr8}
P_{\!_{^{^+}}}^{^{\!_M\!}}\! (t) = -P_{\!_{^{^-}}}^{^{\!_M\!}}\!(t)
\approx
P_{e}^{^{_{_0}}} [1 - e_{}^{-w_n^{} t}\phi (t)]e_{}^{-w_p t},
\end{equation}
where $\, \phi (t) = \cos (2Qt) + (w_n^{}/2Q) \sin (2Qt), $
\begin{equation} \label{spepr9}
P_{e}^{^{_{_0}}} = 4(-1)^{^{\!_S}}\!({}_{\!}J_e^{}Q/|\Omega|^2), \;\mbox{and}
\;\: w_{\!p}^{} = 4w_n^{} (J_e^{2}/|\Omega|^2)
\end{equation}
with $|\Omega|^2 = w_n^2 + 4Q^2$.

In the case $|J_e^{}| \ll  w_{^{\!_\pm}}^{_n}$ formula
(\ref{spepr8}) describes the generation of the CIDEP
$P_{\!_{^{^\pm}}}^{^{\!_M\!}}\! (t)$ for $t \lesssim w_{n}^{-1}$,
the attainment of the maximum amplitude $P_{e}^{^{_{_0}}}$ at $t >
w_{n}^{-1}$, and then slow relaxation with the rate $w_{p}^{} \ll
w_{n}^{}$. The relaxation (induced by the exchange interaction) is
the manifestation of the fact that the observable operators
($S_{_{\pm_z}}$) do not commute with the spin-Hamiltonian $H$.

\subsection{Effect of spin relaxation and reaction}

In the above analysis we have considered the $ST_0^{}$-CIDEP
generation kinetics at relatively short times $t \sim
1/w_{^{\!_\pm}}^{_n} \ll 1/w_{^{\!_\pm}}^{_d}, 1/k_s^{}$. At longer
times the dependence of the $ST_0^{}$-CIDEP
$P_{\!_{^{^\pm}}}^{e\,}\! (t)$ can be strongly affected by slow
population relaxation and recombination. The manifestation of this
effect can be described within the extended model, taking into
account [apart from recombination (\ref{gform5})] relaxation
transitions between all states $|S\rangle,|T_{\pm,0}^{}\rangle$, in
addition to $ST_0^{}$ transitions, considered above.

In the model the relaxation-affected-CIDEP generation kinetics is
described by the extended vector of spin-density matrix elements $
\rho_{{e}}^{} = [\rho_{_{11}}, \rho_{_{22}}, \rho_{_{23}},
\rho_{_{32}}, \rho_{_{33}}, \rho_{_{44}}]_{}^{^{_{\sf T}}}$,
satisfying the SLE of type of eq. (\ref{spepr5})
\begin{equation} \label{spepr10}
(\epsilon + \hat L_{e}^{})\tilde {\rho}_{{e}}^{} = \rho_i^{_0},
\;\;\mbox{with}\;\; \hat L_{e}^{} =
\hat L_{r}^{} + \hat W_{r}^{} + \hat K_s^{e},
\end{equation}
in which $\hat W_{r}^{}$ is the intraparticle relaxation supermatrix
defined by eq. (\ref{gform3}), and $\hat K_s^{e}$ is the part of the
reaction supermatrix (\ref{gform5}), operating in the subspace of
matrix elements $\rho_{{e}}^{}$. The approximate is quite accurate
in the considered strong magnetic field limit, i.e. the limit of
large Zeeman frequences $\omega_{\!_\pm} \gg w_{_{\!\pm}}^{n}$.

Solution of the SLE (\ref{spepr10}) is, in general, fairly
cumbersome and not suitable for the analytical analysis. To simplify
the problem we consider the case of weak reactivity $k_s^{} \ll
w_{_{\!^\pm}}^{_d} \; (\ll w_{_{\!^\pm}}^{_n})$ and weak exchange
interaction, when $w_{\!p} \ll w_{_{\!^\pm}}^{_d}$ [see eq.
(\ref{spepr9})].

In this limit we get the following expression for the Laplace
transform of the difference $\tilde \rho_{_{22\!}} (\epsilon) -
\tilde \rho_{_{33}\!} (\epsilon)$, which determines the
relaxation-affected multiplet CIDEP
\begin{eqnarray} \label{spepr10bb}
\tilde \rho_{_{M}}\! (\epsilon) &=& \tilde \rho_{_{33\!}} (\epsilon)
- \tilde \rho_{_{22}\!} (\epsilon)\nonumber \\
&\approx& (2J_e^{}Q/|\Omega|^2)[\sigma_{\!_S}  - k_s^{}(\epsilon +
k_s^{})^{^{_{-1}}}]D_{\epsilon}^{^{_{-1}}}\!, \qquad
\end{eqnarray}
where $\sigma_{\!_S }= (-1)^{^{\!_S}}\!$ (with $S = 0,1$),
$|\Omega|^2 = w_{\!n}^2 + 4Q^2$, and
\begin{equation} \label{spepr10c}
D_{\epsilon}^{} \!\approx
(\epsilon + 2w_{_{\!+}}^{_d})(\epsilon +
2w_{_{\!-}}^{_{d}})/(\epsilon + w_{\!d}^{})
\end{equation}
with $w_{\!d}^{} \!= w_{_{\!+}}^{_d} + w_{_{\!-}}^{_d}$ and
$w_{\!_n}\! = w_{_{\!+}}^{_n} + w_{_{\!-}}^{_n}.$ Taking into
account the relation $\tilde P_{\!_{^{^+}}}^{^{_M\!}}\! (\epsilon) =
-\tilde P_{\!_{^{^-}}}^{^{_M\!}}\! (\epsilon) = 2\tilde
\rho_{_{M}}\! (\epsilon)$ and making the inverse Laplace
transformation we obtain formula
\begin{equation} \label{spepr10d}
P_{\!_{^{^+}}\!}^{^{_M\!}\!}
(t) =\! - P_{\!_{^{^-}}\!}^{^{_M\!}\!} (t) \approx
P_{e}^{^{_{_0}}\!}[\bar p_{e\!}^{\!} (t) - (_{\!}-_{\!}1_{\!})^{^{\!_S}}\!
(\kappa_s^{}/\bar w_{\!_d})e^{-\kappa_{\!s}^{}t}],\!
\end{equation}
in which $\bar p_{e\!}^{\!} (t) =
\frac{1}{2}\big(e^{-2w_{^{\!+}}^{_d}t} +
e^{-2w_{^{\!-}}^{_d}t}\big)\:$, $\bar w_{\!_d} \approx 4
w_{_{\!+}}^{_d}w_{_{\!-}}^{_d}/w_{\!_d} ,\:$ $\,\kappa_s^{} =
\frac{1}{4}k_s^{}, \,$ and $P_{e}^{^{_{_0}}}\!$ is defined by eq.
(\ref{spepr9}),

The population relaxation results in marked population of states
$|1\rangle$ and $|4\rangle$. It is easily seen that, in general,
this population leads to the net CIDEP $P_{\!_{^{^+}}}^{^{_N\!}}\!
(t)$. The above-discussed solution of the SLE (\ref{spepr10}) yields
simple expression for this (Laplace transformed) net contribution
$\tilde P_{\!_{^{^+}}}^{^{_N\!}}\! (\epsilon) = \tilde
P_{\!_{^{^-}}}^{^{_N\!}}\! (\epsilon) = 2\tilde \rho_{_{\!N}}\!
(\epsilon)$, which appears to be directly related to the multiplet
CIDEP, i.e. to $\tilde \rho_{_{M}} (\epsilon)$:
\begin{equation} \label{spepr10e}
\tilde \rho_{_{\!N}} (\epsilon) = \tilde \rho_{_{44\!}} (\epsilon)
- \tilde \rho_{_{11}\!} (\epsilon) =
[(w_{^{\!_+}}^{_d} - w_{^{\!_-}}^{_d})/w_d^{}]
\tilde \rho_{_{M}} (\epsilon),
\end{equation}
By summing up both multiplet and net contributions we obtain
following formula for the total P${}^{^{_\pm}}_{}\!$-CIDEP
\begin{equation} \label{spepr10f}
P_{\!_{^{\pm}}}^{{e\!}}\!
(t) = (w_{^{\!_\mp}}^{_d}/w_d^{})P_{\!_{^{\pm}}}^{^{_M\!}}\!(t)
\;\;\mbox{with}\;\; w_d^{} = w_{^{\!_+}}^{_d} + w_{^{\!_-}}^{_d}.
\end{equation}
This formula shows that the difference in population relaxation
rates leads to the difference in values of the total
relaxation-affected $ST_0^{}$-CIDEP of P${}^{^{\!_\pm}}_{}\!$
particles (without change of signs).

The expressions (\ref{spepr10d}) and (\ref{spepr10f}) predict
another interesting effect of population relaxation and
spin-selective reactivity: the change of the CIDEP sign for the
singlet initial P${}^{^{\!_+}}_{}\!$P${}^{^{\!_-}}_{}\!$ state (i.e.
for $S = 0$) at certain time $\tau_s^{} $, which can be evaluated by
$P_{\!_{^{^\pm}}}^{^{_M\!}}\!(\tau_s^{}) = 0$ (recall, we assume
that $k_s^{} \ll w_{_{\!\pm}}^{_d}$). In the simplest case
$w_{^{\!_+}}^{_d} \approx w_{^{\!_-}}^{_d}$:
\begin{equation} \label{spepr10g}
\tau_s^{} \approx w_{d\!}^{-1}\! \ln (w_{\!d}^{}/k_s^{}).
\end{equation}

This CIDEP-sign change can easily be understood, taking into account
that at long times the population relaxation leads to almost
complete equilibration of spin subsystem. In this limit the
spin-correlation is generated by recombination in the singlet state.
The recombination-induced decay of the singlet component of
population manifests itself similarly to the case of overpopulation
of the (equilibrated) triplet state [see eq. (\ref{gform5b})], i.e.
the sign of the initial $ST_0^{}$-CIDEP, corresponding to $\rho_i =
\rho_{_{S}}$, is changed by the opposite one (for $\rho_i =
\rho_{_{T}}$).

Note that the proposed mechanism of the change of
population-relaxation-affected $ST_0^{}$-CIDEP sign also predicts
the change of the APS-sign at times $t > \tau_s^{}$ because the
APS-sign depends on the P${}^{^{\!_+}}_{}\!$P${}^{^{\!_-}}_{}\!$
spin coherence as well [see eq. (\ref{spepr3})]:\cite{Hore0} in
other words, the found change of the singlet spin state by the
triplet one at $t > \tau_s^{}$ is accompanied by the change of the
APS-sign.

In the consideration of the relaxation/recombination effect we have
assumed that the reactivity is small $k_s^{} \ll
w_{_{\!^\pm}}^{_d}$. The numerical analysis shows, however, that
this effect persists for $k_s^{} \gtrsim w_{_{\!^\pm}}^{_d}$ as
well.

\section{Numerical analysis of contributions}

\subsection{General remarks}

It is worth beginning the numerical analysis of specific features of
trESR spectra with some general illustrative calculations of APS and
$ST_0^{}$-CIDEP contributions, considered above.

The shape of trESR spectra and, in particular, the amplitudes of
these two contributions are essentially determined by spin-evolution
during the measurement [apart from the spin density matrix
$\rho_t^{}$ at time $t$ of the measurement], i.e. by characteristic
properties of the superoperators $\hat L_{_{^\omega}}^{}$ and $\hat
F_{\!_{^\omega}}^{}$ [see eqs. (\ref{gform6}) and (\ref{gform8})].
General methods of rigorous, but cumbersome calculation of the
contributions have already been discussed in literature.\cite{Shu2}

In the considered limit of weak exchange interaction $J_e^{} \ll
w_g^{},w_{\!^{_\pm}}^{_n},|Q|$, however, the calculation of
contributions can be simplified using the linear approximation in
$\hat J = J_e^{} (S_{}^{2} - 1)$, in which
\begin{equation} \label{numepr1}
\hat F_{\!_{^\omega}}^{} \approx \hat F_{\!_{^\omega}}^{0} +
\hat F_{\!_{^\omega}}^{1}
\;\; \mbox{with}\;\; \hat F_{\!_{^\omega}}^{1} =
\hat F_{\!_{^\omega}}^{0}(i\hat J/w_g^{})\hat F_{\!_{^\omega}}^{0}
\end{equation}
and
\begin{equation} \label{numepr2}
\hat F_{\!_{^\omega}}^{0} = w_g^{}(w_g^{} +
\hat L_{\!_{^\omega}}^{0})_{}^{-1}.
\end{equation}
In formula (\ref{numepr2}) the superoperator $\hat
L_{\!_{^\omega}}^{0} $ is defined by eq. (\ref{gform6}) with $\hat
H_{_{^\omega}} $ replaced by $\hat H_{_{^\omega}}^{0} =
({\omega}-\omega_0^{})\hat S_{_z} + Q (\hat S_{a_z}\!\! - \hat
S_{b_z}).$

Substitution of the expansion (\ref{numepr1}) into eq.
(\ref{gform7}) yields the representation for the trESR spectrum
\begin{equation} \label{numepr3}
I_{t}^{} (\omega) \approx I_{t}^{0} (\omega) + I_{t}^{1} (\omega)
\;\:\mbox{with}\;\: I_{t}^{j} (\omega) =
{\rm Tr} [S_{y'}^{}(\hat F_{\!_{^\omega}}^{j}\rho_t^{})]
\end{equation}
and $j = 1,2$. This formula is found to be very accurate (with
relative error $\delta < 10^{-2}$) for parameters of the model used
in the analysis.

Previous investigation\cite{Shu2} and the analysis of Sec. IV.A show
that the term $I_{t}^{0} (\omega)$, in which the effect of $\hat J$
on measurement is neglected, describes the relaxation-affected
$ST_0^{}$-CIDEP (multiplet and net). The term $I_{t}^{1} (\omega)$,
taking into account $\hat J$-effect in linear in $\hat J$
approximation, represents APS\cite{Shu2} (see also the analysis in
Sec. IV.A).

{\it a. APS-contribution.} For small dephasing rates
$w_{^{^\pm}}^{_n} \ll |Q|$ the APS is of the (conventional) shape of
two E/A lines, located at ESR frequences $\omega_{_\pm}$ of
P${}^{^{\!_\pm}}_{}\!$ polarons (in accordance with the analysis in
Sec. IV.A). For $w_{\pm}^{_n} \gtrsim |Q|$, however, the APS shape
drastically changes:\cite{Shu2} with the increase of rates
$w_{^{_\pm}}^{_n}$ the two E/A lines collapse into one of A/E shape
at the center of the spectrum.

{\it b. CIDEP-contribution.} The shape of the (extended $ST_0^{}$)
CIDEP contribution $I_{t}^{0}(b)\equiv I_{t}^{0} (\omega(b))$
remains nearly the same for any relation between
$w_{\!^{_\pm}}^{_n}$ and $ |Q|$.

{\it c. Relaxation/recombination effect.} Predicted effect of the
relaxation/recombination induced change of signs of APS and
$ST_0^{}$-CIDEP contributions is completely confirmed by the exact
numerical calculations in the considered symmetric variant of the
system. The analysis demonstrates the sign change for $w_{\pm}^{_n}
\sim Q$. In the calculation we deliberately consider the case
$k_s^{} \sim w_{_{\!\pm}}^{_d}$ to demonstrate the occurrence of the
sign change independently of the relation of between $k_s^{}$ and
$w_{_{\!\pm}}^{_d}$ (not only for $k_s^{} \ll w_{_{\!\pm}}^{_d}$, as
it is proved in Sec. IV.A.).

\subsection{Analysis of experimental spectra}

Formulas (\ref{numepr1}) - (\ref{numepr3}) allow for the
representation of calculated spectra, reproducing the experimental
results, as a sum of APS and $ST_0^{}$-CIDEP contributions. Figure 2
shows this representation for two times. Noteworthy is that both
contributions are somewhat distorted by the temperature effect
(effect of $p_{_\pm}^{} = e^{-\omega_{\pm}^{}/(k_{B}T)}  < 1$),
resulting in the conventional thermal contribution to trESR spectra.
This distortion manifests itself in the $ST_0^{}$-CIDEP contribution
at long times $t \gtrsim 1/w_{\!_{^\pm}}^{_d}$.

The presented results clearly demonstrate that the observed
significant time dependence of the shape of ESR spectra results from
the subtle superposition of APS and $ST_0^{}$-CIDEP contributions.

1) In particular, at early time $t = 0.8 \, {\rm \mu s}$ the trESR
spectrum looks like the APS-type one, as was mentioned in ref. [24].
The proposed numerical analysis shows, however, that this spectrum
is a sum of APS and $ST_0^{}$-CIDEP contributions of about
comparable amplitude (Fig. 2). Moreover, despite the conventional
APS-like shape of the experimental spectrum, actual numerically
evaluated APS contribution to this spectrum is of the sign opposite
to that for the conventional APS, because of fast dephasing rates
$w_{_\pm}^{_n} \sim |Q|$ (see Sec. IV.A.). As for the main E/A
component of the experimental spectrum, it is, in fact, determined
by the $ST_0^{}$-CIDEP contribution.

\begin{figure}
\setlength{\unitlength}{1cm}
\includegraphics[height=6.2cm,width=6.8cm]{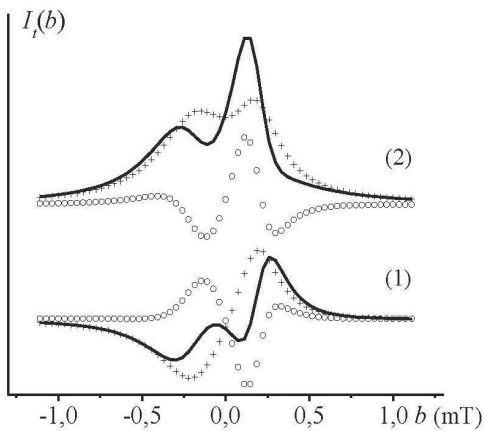}
\caption{Representation of trESR spectra $I_{t}^{}(b)$ (full lines),
numerically calculated for (1) $t = 0.8\, {\rm \mu s}$ and (2) $ t =
3.5\, {\rm \mu s}$ (see Fig. 1), as sums of two contributions
$I_t^{0} (b)$ ($+$) and $I_t^{1} (b)$ (${\tt o} $) [eq.
(\ref{numepr3})] (other parameters are the same as in Fig. 1).}
\end{figure}

The comparable APS and $ST_0^{}$-CIDEP contributions are observed at
longer times as well, manifesting themselves in the asymmetric
distortion of trESR spectra, distinguishable even at fairly long
times $t \sim 10 \, {\rm \mu s}$.

2) Numerical results, displayed in Fig. 2, also demonstrate the
change of the sign of these contributions with the increase of time.
It is seen that at $t = 0.8 \, {\rm \mu s}$ the APS sign is opposite
to that at $t = 3.5 \, {\rm \mu s}$. The same is valid for
$ST_0^{}$-CIDEP contribution, which shows itself in the E/A
distortion of the trESR spectra at short time $t = 0.8 \, {\rm \mu
s}$ changed by A/E one at longer times $t > 3 \, {\rm \mu s}$
(superimposed on the thermal ESR signal), thus confirming the
theoretical prediction Sec. IV.A.

\section{Conclusions}

This work concerns the analysis of specific features of experimental
trESR spectra of photoexcited spin correlated polaron pairs
(P${}^{^{\!_+}}_{}\!$P${}^{^{\!_-}}_{}\!$) in polymer:fullerene
(P3HT-PCMB) blend\cite{Beh1} at low temperature $100\,$K (see Sec.
I). The analysis is mainly based on exact numerical solution of the
SLE within the model of the exponentially decaying polaron pairs,
spin-evolution of which is described by the spin-Hamiltonian
(\ref{gform2}), spin-selective reactivity (\ref{gform5}), and spin
relaxation of two types: free polaron relaxation (\ref{gform3}) and
spin-exchange-induced one (\ref{gform5a}).

According to the analysis, in
P${}^{^{\!_+}}_{}\!$P${}^{^{\!_-}}_{}\!\!$-pairs (in
polymer:fullerene blend under study) the spin dephasing is much
faster than the population relaxation. This relation allows for
approximate analytical analysis of trESR spectra, providing deep
insight into distinctive properties of the time dependence of
spectra. These properties are found to be determined by the specific
features of antiphase (APS) and population-relaxation affected
$ST_0^{}$-CIDEP contributions. The combined numerical/analytical
analysis shows that the generation of the contributions is governed
by sophisticated interplay of coherent spin evolution and
spin-relaxation/recombination of
P${}^{^{\!_+}}_{}\!$P${}^{^{\!_-}}_{}\!$ pairs.

It is found, in particular, that the observed short time trESR
spectra (of the shape of conventional APS with large E/A components)
are, actually, the sums of APS and $ST_0^{}$-CIDEP contributions of
comparable amplitude. As for the long time spectra, they are
determined by the population-relaxation affected spin-selective
recombination of P${}^{^{\!_+}}_{}\!$P${}^{^{\!_-}}_{}\!$ pairs. The
contribution of the relaxation/recombination mechanism dominates at
times longer than the time $ \tau_s^{}$, defined in eq.
(\ref{spepr10g}). For the singlet initial condition this mechanism
predicts the change of the sign of both contributions with time,
distinctively manifesting itself in trESR spectra.

The analysis also shows that the dissociation of
P${}^{^{\!_+}}_{}\!$P${}^{^{\!_-}}_{}\!$ pairs does not strongly
affect the trESR spectra of the system under study at relatively
short times $t < 10\, {\rm \mu s}$ (of main interest in our work).
For this reason, in the calculation we have neglected the effect of
dissociation.  At longer times the contribution of dissociation to
trESR spectra becomes important, but, unfortunately, the accuracy of
available experimental spectra at these times is not high enough for
accurate extraction of the contribution of dissociated pairs to the
observed spectra.

At the end of our work it is worth discussing some recent numerical
studies of the processes similar to that studied here. As an
example, let us discuss one of the most recent and fairly detailed
work, concerning the analysis of trESR spectra of geminate pairs of
neutral radicals in micelles.\cite{Forb7} Analysis is carried out
within the model, which takes into account the spin evolution,
accompanied by relative diffusion. The trESR spectra, investigated
in this work, do not demonstrate the change of the sign of
distortions (of APS and CIDEP-type) at long times. This is,
probably, because the radical pairs are created in the triplet spin
state.

Certainly, the results of the analysis, presented in ref. [18], are
very interesting and insightful. Unfortunately they can hardly be
directly applied to the theoretical investigation of considered spin
correlated P${}^{^{\!_+}}_{}\!$P${}^{^{\!_-}}_{}\!\!$-pairs in
organic solids at low temperatures, first of all because for these
systems the diffusion approximation does not seem to be valid,
especially taking into consideration strong Coulomb interaction
between polarons in non-polar organic solids, the effect of which is
expected to be very important at low temperatures.

Noteworthy is that the direct comparison of parameters of our model
with those of model used in the work [18] is hardly possible. It
worth also adding, that the proposed decomposition of the trESR
spectra into the APS and CIDEP contributions (\ref{numepr3}) cannot
be easily realized in the model proposed in [18].

Concluding our short discussion we would like to emphasize that in
spite of great attention, payed to the interpretation of
experimental trESR spectra of particular spin-correlated
P${}^{^{\!_+}}_{}\!$P${}^{^{\!_-}}_{}\!$ pairs, the goal of our
treatment is more general. The main idea is to develop some new
combined numerical/theoretical methods of the analysis of these
spectra to get some additional insight into the specific features of
P${}^{^{\!_+}}_{}\!$P${}^{^{\!_-}}_{}\!\!$-spin evolution. We belive
that the proposed analysis is fairly convincing illustrative example
of application of the proposed methods.

\textbf{Acknowledgements.}\, The work was partially supported by the
Russian Foundation for Basic Research (project 16-03-00052).

\end{document}